# A mass conserved reaction-diffusion system captures properties of cell polarity

Mikiya Otsuji,[a,*] Shuji Ishihara,[b] Carl Co,[c] Kozo Kaibuchi,[d] Atsushi Mochizuki,[b] and Shinya Kuroda[e]

[a]Department of Anesthesiology, Faculty of Medicine, University of Tokyo, 7-3-1 Hongo, Bunkyo-ku, Tokyo 113-8655, Japan

[b]Division of Theoretical Biology, National Institute for Basic Biology, 38 Nishigonaka, Myodaiji, Okazaki, Aichi, 444-8585, Japan

[c]Department of Cellular and Molecular Pharmacology and Program in Biological Sciences, University of California, San Francisco, San Francisco, CA 94143, USA

[d]Department of Cell Pharmacology, Graduate School of Medicine, Nagoya University, 67 Tsurumai, Showa, Nagoya, Aichi, 466-8550, Japan

[e]Department of Biophysics and Biochemistry, Graduate school of Science, University of Tokyo, 7-3-1 Hongo, Bunkyo-ku, Tokyo 113-0033, Japan

Corresponding author: Mikiya Otsuji, Department of Anesthesiology, Faculty of Medicine, University of Tokyo, 7-3-1 Hongo, Bunkyo-ku, Tokyo 113-8655, Japan

Tel.: +81-3-5841-4697

FAX: +81-3-5841-4698

E-mail: mohtsuji-tky@umin.ac.jp

Abbreviations: GDP/GTP, guanosine 5'-bis/tri phosphate; PI3K, phosphoinositide 3-kinase; PTEN, phosphatase and tensin homologue deleted on chromosome 10; $PIP_3$, phosphatidylinositol 3,4,5-triphosphate; GEF, guanine nucleotide exchange factor; GAP, GTPase-activating proteins.

**Abstract**

Various molecules exclusively accumulate at the front or back of migrating eukaryotic cells in response to a shallow gradient of extracellular signals. Directional sensing and signal amplification highlight the essential properties in the migrating cells, known as cell polarity. In addition to these, such properties of cell polarity involve unique determination of migrating direction (uniqueness of axis) and localized gradient sensing at the front edge (localization of sensitivity), both of which may be required for smooth migration. Here we provide the mass conservation system based on the reaction-diffusion system with two components, where the mass of the two components is always conserved. Using two models belonging to this mass conservation system, we demonstrate through both numerical simulation and analytical approximations that the spatial pattern with a single peak (uniqueness of axis) can be generally observed and that the existent peak senses a gradient of parameters at the peak position, which guides the movement of the peak. We extended this system with multiple components, and we developed a multiple-component model in which cross-talk between members of the Rho family of small GTPases is involved. This model also exhibits the essential properties of the two models with two components. Thus, the mass conservation system shows properties similar to those of cell polarity, such as uniqueness of axis and localization of sensitivity, in addition to directional sensing and signal amplification.

## 1. Introduction

Eukaryotic cells, such as *Dictyostelium* and neutrophils, respond to temporal and spatial gradients of extracellular signals with directional movements (Affolter and Weijer, 2005; Chung, et al., 2001; Comer and Parent, 2002; Iijima et al., 2002; Parent and Devreotes, 1999; Ridley et al., 2003). This process, known as chemotaxis, is one of the fundamental properties of cells (Carlos, 2001; Firtel and Meili, 2000; Gillitzer and Goebeler, 2001). In a migrating cell, specific molecular events take place at the front edge (anterior) and at the back (posterior) (Affolter and Weijer, 2005; Chung et al., 2001; Iijima et al., 2002; Xu et al., 2003). The spatially distinctive molecular accumulation inside cells is known as cell polarity. The front-back polarity usually has one axis, and this uniqueness is an important property because a migrating cell with two fronts could not move effectively (Foxman et al., 1997). Another reported property of the front-back polarity is higher sensitivity of the front to a gradient of extracellular signals (Devreotes and Janetopoulos, 2003; Xu et al., 2003). This property would also be important because the direction of travel should be controlled at the front edge.

Many mathematical models that account for gradient sensing and signal amplification in cell polarity have been proposed (e.g., Devreotes and Janetopoulos, 2003). The local excitation and global inhibition model has been proposed to explain spatial gradient sensing (Ma et al., 2004; Parent and Devreotes, 1999). Some models involve positive feedback loops for amplified accumulation of signaling molecules (Levchenko and Iglesias, 2002; Narang et al., 2001; Postma and Van Haastert, 2001; Skupsky et al., 2005; Subramanian and Narang, 2004). A reaction-diffusion model that includes local self-enhancement and long-range antagonistic effects has been proposed for directional sensitivity (Meinhardt, 1999). These models explain gradient sensing and signal amplification well enough, but it is still not clear what mechanism gives cell polarity the unique axis and the localized sensitivity. In this study, we focus on these two properties and describe a new mechanism.

Many biological phenomena have been explained by reaction-diffusion systems. One of the most famous and extensively studied reaction-diffusion models is the Turing model, in which robust spatial patterns, such as stripes or spots, emerge via a diffusion-driven instability of the homogenous stationary states (Kondo, 2002; Meinhardt and Gierer, 2000). An ordinary Turing pattern in one-dimensional space is stripes with intrinsic scale length (Murray, 2003), and some studies of reaction-diffusion systems have reported proportional behavior for the system sizes (Ishihara and Kaneko, 2006; Othmer and Pate, 1980). But there are few reaction-diffusion models that can explain the uniqueness of concentration peak in the system.

In this study, we found that some reaction-diffusion systems with mass conservation of components always exhibited specific accumulation of components at a single site, independent of the system size. Because these mass conservation systems can explain the unique concentration peak, we analyze two mass conservation models by numerical simulations and one model by analytical approximations. We demonstrate that models have one-peak stationary states regardless of the system sizes and that multiple-peak stationary states are unstable. To investigate how the systems detect the gradient of signals, whose levels vary with location, we also analyze mass conservation models with position-dependent parameters that are described by functions of position. We show that an existent peak moves depending on the gradient of the parameter value and that the velocities are decided at the site of the peak.

The Rho family of small GTPases have been reported to transit between two states, an active and an inactive state (Kaibuchi et al., 1999). These two states can be treated as components of the mass conservation model. Here we proposed a model based on cross-talk of Rho GTPases and show that this model also exhibits the polarity with one axis.

In this paper, we introduce two mass conservation models and analyze the models by numerical simulations and analytical approximations. Next, we describe the Rho GTPases model, and we discuss the mass conservation system in terms of biological validity.

## 2. Models

We investigate the following class of reaction-diffusion system composed of two components, u and v:

$$\frac{\partial u}{\partial t} = D_u \frac{\partial^2 u}{\partial x^2} - f(u,v), \tag{1a}$$

$$\frac{\partial v}{\partial t} = D_v \frac{\partial^2 v}{\partial x^2} + f(u,v), \tag{1b}$$

$$D_u < D_v, \tag{1c}$$

where $u$ and $v$ denote the concentrations of u and v, respectively, at time $t$ and at position $x$, and $D_u$ and $D_v$ denote the diffusion coefficients of u and v, respectively. Reaction terms $f$ are given by the function of $u$ and $v$. Because the total quantity of u and v is conserved in this case, we refer to this system as the "mass conservation system."

To investigate the features of mass conservation systems, we propose two models, Model I and Model II, described as follow:

Model I

$$f(u,v) = a_1(v - a_2 u)\left(v - \frac{a_3}{u+v}\right),$$
$$D_u < D_v,$$
$$a_1 = 0.2,\ a_2 = 2,\ a_3 = 2,\ D_u = 0.01,\ D_v = 1, \tag{2}$$

Model II

$$f(u,v) = -a_1(u+v)\left[(\alpha u + v)(u+v) - a_2\right],$$
$$D_u = \alpha D_v \, ,$$
$$\alpha < 1 \, ,$$
$$a_1 = 0.5, \, a_2 = 2.2, \, D_u = 0.1, \, D_v = 1 \, or \, 2,$$

(3)

where $a_1$, $a_2$, $a_3$, and $\alpha$ are parameters of the models.

We show the results of numerical simulations using Model I in Section 3 and those of analytical approximation using Model II in Section 4. We consider a one-dimensional circular system with circumference $L$. The position is represented by $x$ ($-L/2 \leq x \leq L/2$). We apply the periodic boundary condition, which is used in many models explaining cell polarity (Meinhardt, 1999; Subramanian and Narang, 2004).

## 3. Results of numerical simulations

In this section, we demonstrate the characteristic behavior observed in the numerical simulations of the mass conservation model. We only present the results of Model I, but Models I and II exhibit qualitatively similar properties, at least for the properties discussed in this paper. Simulations were performed using explicit difference methods. The difference intervals for calculations of Model I were taken to be $\Delta t = 0.005$ and $\Delta x = 0.2$.

### 3.1 Transient behavior and final stable states of the system

We observed the transition of the system from the homogenous stationary state with small perturbations. The reaction terms and diffusion coefficients of Eqs. (1a–c) were given by Eqs. (2) (Model I). The initial state was given by $u = 1 + 0.001Rnd$ and $v = 1 + 0.001Rnd$, where $Rnd$ values are random numbers ($-1 < Rnd < 1$), representing spatial perturbations. Note that this initial state satisfies $f(u,v) = 0$. The system sizes were taken as $L = 10$ or 20. The results of simulations are shown in Fig. 1A ($L = 10$) and Fig. 1B ($L = 20$). The systems transiently formed wave patterns with multiple peaks, but progressively lost their peaks and finally stabilized with one peak. We show the transitions of the heights of four peaks generated in the system with $L = 20$ (Fig. 1C). Two peaks vanished in a short time and one of the surviving peaks also disappeared after a certain period of time. The number of peaks did not increase after the initial wave pattern was generated.

We can understand the earliest phase of these transitions by linearization analyses around the homogenous states. In the homogenous stationary state, the Jacobian matrix for the reaction terms is given by

$$\mathbf{J} = \begin{pmatrix} -f_u & -f_v \\ f_u & f_v \end{pmatrix}, \tag{4}$$

where $f_u$ and $f_v$ denotes the partial derivatives of $f$ by $u$ and $v$, respectively, at a homogenous stationary state of Eqs. (1a–c). Through the stability analysis using $\mathbf{J}$, the range of wave numbers ($k_h$) that have positive eigenvalues is obtained as $0 < k_h < \left( \dfrac{D_u f_v - D_v f_u}{D_u D_v} \right)^{\frac{1}{2}}$, and the wave number that has the largest eigenvalue and grows most rapidly from the homogenous state, $k_h^*$, is obtained as follows:

$$k_h^* = \left\{ \frac{1}{D_v - D_u} \left[ (f_u + f_v) + \left( \sqrt{\frac{D_u}{D_v}} + \sqrt{\frac{D_v}{D_u}} \right) \sqrt{f_u f_v} \right] \right\}^{\frac{1}{2}}. \tag{5}$$

For Model I (Eqs. 2), we obtain $k_h^* = 1.17$; the most growable wavelength is $2\pi / k_h^* = 5.38$. Therefore, the system with $L = 10$ generates two peaks and the system with $L = 20$ generates four peaks from the homogenous states.

Repeated computations always showed a final one-peak pattern, regardless of the system sizes ($L = 5, 10, 20, 40$; see Discussion). To confirm that the coexistence of multiple peaks is unstable, we performed the following experimental simulations. First, we obtained a stable one-peak pattern in Model I (Eqs. 2) by setting $L = 5$ and taking the initial state as $u = 1$ and $v = 1$. Because we applied the periodic boundary condition to this system, we could set the center of the concentration peak at $x = 0$ by translation. Next, by duplicating this profile ($L = 5$) and connecting them, we obtained a new profile ($L$ = 10) with two peaks. We used this profile ($L$ = 10) with small perturbations (+0.01$Rnd$) as the initial state of the following simulation. As shown in Fig. 1D, one of the peaks collapsed with time, and only one peak persisted.

### 3.2 Behavior of the system including a position-dependent parameter

To investigate the behavior of the system in response to the environment, which is represented as a parameter in the reaction terms, we observed the transition of the system with a position-dependent parameter. We substituted $a_3^* = 2 + 0.06\sin\left(2\pi\dfrac{x}{L}\right)$ for $a_3 = 2$ in Eqs. (2). The system size was given by $L$ = 10 and the initial state was given by $u$ = 1 and $v$ = 1. The result of this simulation is shown in Fig. 2A. One peak arose at $x$ = -0.25, where the value of $a_3^*$ was smallest in the system, and the peak was stabilized.

Next, we observed how an existent peak responded to a new position-dependence in parameter $a_3$. We obtained a stable one-peak pattern in Model I (Eqs. 2) by setting $L$ = 10 and taking the initial state as $u$ = 1 and $v$ = 1. We set the center of the concentration peak at $x$ = 0 by translation, and we substituted $a_3^* = 2 + 0.06\sin\left(2\pi\dfrac{x}{L}\right)$ for $a_3 = 2$ in Eqs. (2). The simulation showed the movement of the concentration peak toward the negative $x$ direction, where the value of $a_3^*$ was smaller. This result indicates that the system is sensitive to the position-dependence of a parameter.

To examine whether the sensitivity is localized, we observed the movement of an existing peak responding to local position-dependence of $a_3$. We obtained stable one-peak patterns in Model I (Eqs. 2) by setting $L$ = 10 and taking the initial state as $u$ = 1 and $v$ = 1. We set the center of the concentration peak at $x$ = -1.4 or $x$ = -3.4 by translation. Then we substituted the following

$a_3^*$ for $a_3 = 2$ in Eqs. (2):

$$a_3^* = \begin{cases} 2 & \left( \dfrac{x}{L} < -0.5, \dfrac{x}{L} > 0.5 \right) \\ 2 - 0.08\cos\left(2\pi\dfrac{x}{L}\right) & \left( -0.5 \leq \dfrac{x}{L} \leq 0.5 \right) \end{cases} . \qquad (6)$$

The results of the simulations are shown in Fig. 2C (peak at $x$ = -1.4) and Fig. 2D (peak at $x$ = -3.4). In the case where there was overlap between the existent peak and the local position-dependence of $a_3^*$, the peak moved according to the gradient of $a_3^*$. In the case where there was little overlap, however, the peak hardly moved. These results indicate that the sensitivity to the position-dependence of a parameter is localized at the site of the peak.

In this paper, we chose $a_3$ as a position-dependent parameter, but similar results were observed when we chose $a_1$ or $a_2$ (data not shown).

## 4. Analysis of Model II

To better understand the results of the numerical simulations, we investigate Model II (Eqs. 3) by analytical approximations. Here we show that: (1) the model has one-peak stationary states, regardless of the system size (if not too small); (2) multiple-peak stationary states are unstable; and (3) the existent peak moves depending on the gradient of the parameter value, and the sensitivity is localized. Finally, we verify our analyses by comparing analytical results with the values obtained by numerical simulations.

### 4.1 Existence of a one-peak stationary solution

We define the following variables and function:

$$N = u + v ,\tag{7a}$$

$$P = D_u u + D_v v ,\tag{7b}$$

$$f^*(N, P) = f\left(\frac{D_v N - P}{D_v - D_u}, \frac{P - D_u N}{D_v - D_u}\right).\tag{7c}$$

Equations (1a, b) are rewritten as the following set of partial differential equations for $N$ and $P$:

$$\frac{\partial N}{\partial t} = \frac{\partial^2 P}{\partial x^2} ,\tag{8a}$$

$$\frac{\partial P}{\partial t} = (D_u + D_v)\frac{\partial N}{\partial t} - D_u D_v \frac{\partial^2 N}{\partial x^2} + (D_v - D_u)f^*(N, P).\tag{8b}$$

Under the periodic boundary conditions, the stationary solutions of Eqs. (8a, b), $N_e(x)$ and $P_e(x)$, satisfy the following equations:

$$P_e(x) = P_e \quad (uniform), \tag{9a}$$

$$\frac{d^2}{dx^2} N_e(x) = \frac{D_v - D_u}{D_u D_v} f^*\big(N_e(x), P_e\big). \tag{9b}$$

Consider the case of Model II given by Eqs. (3). For any $P_e$ (> 0), Eq. (9b) with substitution of Eq. (3) has a family of periodic solutions with periods between $\lambda_{min} < \lambda < \infty$ (see Appendix). Here $\lambda_{min}$ is given by $\lambda_{min} = 2\pi \sqrt{\dfrac{D_u D_v}{D_v - D_u} \dfrac{1}{a_1 a_2}}$. Thus, for an arbitrary large system with $L > \lambda_{min}$, one can choose a one-peak solution that satisfies the boundary conditions at $x = \pm L/2$ by taking $\lambda = L$. Note that $P_e$ is related to the average mass of $N_e(x)$, $\overline{N} = \dfrac{1}{L}\int_{-L/2}^{L/2} N_e(x)dx$. For a sufficiently large period ($\lambda \to \infty$), this is expressed as

$$P_e \to \frac{6 D_v a_2}{\lambda \overline{N}} \sqrt{\frac{D_u D_v}{D_v - D_u} \frac{1}{a_1 a_2}}. \tag{10}$$

Thus, for a sufficiently large system, $N_e(x)$ and $P_e(x)$ are approximated as:

$$N_e(x) = N_0 \operatorname{sech}^2\big[b(x - x_p)\big], \tag{11a}$$

$$P_e(x) = P_e, \tag{11b}$$

where $x_p$ denotes the center of the peak and $b$, $N_0$, and $P_e$ are constants given by:

$$b = \frac{1}{2}\sqrt{\frac{D_v - D_u}{D_u D_v} a_1 a_2} \; , \tag{12a}$$

$$N_0 = \frac{L\overline{N}b}{2} \, , \tag{12b}$$

$$P_e = \frac{3D_v a_2}{L\overline{N}b} \, . \tag{12c}$$

Here we obtain the one-peak solution for Model II (Eqs. 3) by setting an arbitrary $L$ and $\overline{N}$ .

## 4.2 Stability of periodic solutions

### 4.2.1 Stability of one-peak solution

First, we consider the stability of a one-peak solution given by Eqs. (11a, b). Without loss of generality, we set $x_p = 0$ here. We set $N(x,t) = N_e(x) + \Delta N(x,t)$ and $P(x,t) = P_e + \Delta P(x,t)$, and the stability is estimated by a linearized equation of Eqs. (8a, b) around Eqs. (11a, b), given as follows:

$$\frac{\partial \Delta N}{\partial t} = \frac{\partial^2 \Delta P}{\partial x^2} \, , \tag{13a}$$

$$\frac{\partial \Delta P}{\partial t} = (D_u + D_v)\frac{\partial \Delta N}{\partial t} - D_u D_v \frac{\partial^2 \Delta N}{\partial x^2} + (D_v - D_u)(h_N \Delta N + h_P \Delta P) \, , \tag{13b}$$

where $h_N(x)$ and $h_P(x)$ are partial derivatives of $f^*(N,P)$ by $N$ and $P$, respectively, at the solution Eqs. (11a, b), that is, $h_N(x) = \partial f^*(N_e(x), P_e) / \partial N$ and $h_P(x) = \partial f^*(N_e(x), P_e) / \partial P$. Let us represent $(\Delta N, \Delta P)$ as $\left(e^{\mu t} n_\mu(x), e^{\mu t} p_\mu(x)\right)$ and consider the case of Model II. Equations (13a, b) lead to the following:

$$\frac{d^2 p_\mu}{dx^2} = \mu \, n_\mu \, , \qquad\qquad\qquad\qquad (14a)$$

$$\frac{d^2 n_\mu}{dx^2} = \left\{\frac{D_u + D_v}{D_u D_v} \mu + 4b^2 \left[1 - 3\,\mathrm{sech}^2(bx)\right]\right\} n_\mu - \left[\frac{\mu}{D_u D_v} + \frac{6b^2 N_0}{P_e}\,\mathrm{sech}^4(bx)\right] p_\mu. \quad (14b)$$

If there is nontrivial $\left(n_\mu(x), p_\mu(x)\right)$ that satisfies Eqs. (14a, b) for $\mu$ with a positive real part, the solution is unstable. Note that $(n_{\mu 0}, p_{\mu 0}) = \left(n_0\,\mathrm{sech}^2(bx), -n_0 P_e / N_0\right)$ satisfies Eqs. (14a, b) for $\mu = 0$ under periodic boundary conditions. Here $n_0$ is an arbitrary factor, originated from the linearity of equations, and we can set $n_0 = 1$. For $\mu$ with an absolute value near zero, we can obtain $(n_\mu, p_\mu)$ by the expansion from $(n_{\mu 0}, p_{\mu 0})$ with regard to $\mu$. To do this, we take $n_\mu = n_{\mu 0} + \mu \, n_{\mu 1} + ..$ and $p_\mu = p_{\mu 0} + \mu \, p_{\mu 1} + ...$ In the first order of the expansion, $(n_{\mu 1}, p_{\mu 1})$ obeys the following equations:

$$\frac{d^2 p_{\mu 1}}{dx^2} = \mathrm{sech}^2(bx), \qquad\qquad\qquad\qquad (15a)$$

$$\frac{d^2 n_{\mu 1}}{dx^2} = 4b^2 \left[1 - 3\,\mathrm{sech}^2(bx)\right] n_{\mu 1} - \frac{6b^2 N_0}{P_e}\,\mathrm{sech}^4(bx) p_{\mu 1} + \frac{D_u + D_v}{D_u D_v}\,\mathrm{sech}^2(bx) + \frac{1}{D_u D_v}\frac{P_e}{N_0}. \quad (15b)$$

Thus, $p_\mu$ is immediately derived from Eq. (15a) as:

$$p_\mu(x) = -\frac{P_e}{N_0} + \frac{\mu}{b^2}\left\{\log[\cosh(bx)] + C_1 x + C_2\right\}, \tag{16}$$

and $n_\mu$ is obtained by solving Eq. (15b) with substitution of $p_{\mu 1}$. $C_2$ is determined by the mathematical condition that $n_{\mu 1}$ should be orthogonal to $n_{\mu 0}$. In practice, $C_2$ has little influence on $(n_\mu, p_\mu)$, and we set $C_2 = 0$ in the following analysis. We illustrate $n_\mu$ and $p_\mu$ in Fig. 3A and 3B, respectively. Note that, in the region where $|x|$ is larger than $1/b$, $p_\mu$ can be approximated as follows:

$$\begin{cases} p_\mu \approx -\dfrac{P_e}{N_0} + \mu \dfrac{-b + C_1}{b^2} x & (x < 0) \\ p_\mu \approx -\dfrac{P_e}{N_0} + \mu \dfrac{b + C_1}{b^2} x & (x > 0) \end{cases}. \tag{17}$$

This approximation is verified numerically, as shown by the dashed line with an arrow in Fig. 3B. Considering periodic boundary conditions, $p_\mu(-L/2)$ and $p_\mu(L/2)$ should connect smoothly, therefore, Eq. (16) never satisfies Eqs. (14a, b) except for $\mu = 0$.

### 4.2.2 Instability of multiple-peak solutions

Here we consider the case of multiple-peak solutions. As long as $L/n$ is larger than $1/b$, an identical $n$-peak periodic solution is approximated by:

$$N_e^{\ j}(x) = \frac{1}{n} N_0 \, \mathrm{sech}^2\left(x - x_p^{\ j}\right),$$ (18a)

$$P_e^{\ j}(x) = n P_e \, ,$$ (18b)

on each $j$th domain defined as $x_p^{\ j} - L/2n \leq x \leq x_p^{\ j} + L/2n$, where $x_p^{\ j}$ is the position of each peak$_j$ ($j = 1, 2, ... n$) given by $x_p^{\ j} = -L/2 + \left(j - 1/2\right)L/n$. We set $\Delta P = P(x,t) - P_e^{\ j}(x) = e^{\mu t} \, p_\mu^{\ j}$. As in the case of the one-peak solution, we can get

$$p_\mu^{\ j} = n_0^{\ j} \left\{ -\frac{n^2 P_e}{N_0} + \frac{\mu}{b^2} \left[ \log\left\{ \cosh\left[b\left(x - x_p^{\ j}\right)\right]\right\} + C_1^{\ j}\left(x - x_p^{\ j}\right)\right]\right\},$$ (19)

for each $j$th domain. For multiple-peak solutions, $n_0^{\ j}$ and $C_1^{\ j}$ are determined by the boundary conditions such that $p_\mu^{\ j}$ connect smoothly between adjacent domains. Because Eq. (17) is a good approximation at the respective domain boundaries, it is helpful to consider the boundary conditions; it is enough to consider a piecewise linear function such that it takes the value $p_j = -n_0^{\ j} \, n^2 P_e / N_0$ at $x = x_p^{\ j}$, and at the point the slope changes by $-\mu\left(2N_0 / n^2 P_e b\right) p_j$. A smooth connection between them means the slope is given as $\left(p_{j+1} - p_j\right)/(L/n)$ for $x_p^{\ j} < x < x_p^{\ j+1}$. Thus, $p_j$ satisfies the following relationship:

$$\frac{p_{j+1} - p_j}{L/n} - \frac{p_j - p_{j-1}}{L/n} = -\mu \frac{2N_0}{n^2 P_e b} p_j \, .$$ (20)

Considering periodic boundary conditions, we obtain the following solutions:

$$p_j{}^{(k)} = \cos\left(\frac{2k\pi}{n}j + \theta_0\right), \tag{21a}$$

$$\mu^{(k)} = \frac{2n^3 P_e b}{N_0 L}\sin^2\left(\frac{k\pi}{n}\right) = \frac{12n^3 D_s a_2}{L^3 \overline{N}^2 b}\sin^2\left(\frac{k\pi}{n}\right), \tag{21b}$$

where $k$ is an integer ($1 \le k \le n/2$), corresponding to a mode of perturbation, and $\theta_0$ is an arbitrary constant. All modes of perturbations have positive $\mu$, and therefore can grow. Figures 3C–F show the most growable mode for the $n$-peak solution ($n = 2, 3, 4, 5$), that gives the largest value to $\mu$.

Based on this analysis, the one-peak solution is stable, whereas the multiple-peak solutions are unstable.

### 4.3 Dynamics of an existent peak in response to a position-dependent parameter

Here we study the case in which the system has a position-dependent parameter by substituting $a_2^* = a_2 + \varepsilon\, a_\varepsilon(x)$ for $a_2$, where $\varepsilon a_\varepsilon(x)$ is sufficiently smaller than $a_2$. It is useful to represent the explicit dependence of $f^*$ on the parameter, as $f^*\left(N, P, a_2^*\right)$. Consider that there exists a one-peak stationary solution given by Eqs. (11a, b). Without loss of generality, we set $x_p = 0$ here. By replacing $a_2$ by $a_2^*$, the steady one-peak solution is modulated, as represented by:

$$N(x,t) = N_0\,\mathrm{sech}^2\left[b\left(x - \Delta x_p\right)\right] + n_\varepsilon(x), \tag{22a}$$

$$P(x,t) = P_e + p_\varepsilon(x), \tag{22b}$$

where the first terms indicate the unperturbed solution with its peak at $x = \Delta x_p$ without changing the shape of solution, and second terms represent the modulation of the shape. Here $\Delta x_p$ depends on $t$ for the modulation. When $\varepsilon$ is small, $n_\varepsilon$, $p_\varepsilon$, and $\Delta x_p$ are also small. Therefore, we can rewrite Eq. (22a) as follows:

$$N(x,t) = N_0 \operatorname{sech}^2(bx) + 2\Delta x_p b N_0 \tanh(bx)\operatorname{sech}^2(bx) + n_\varepsilon(x). \qquad (22a')$$

The linearized versions of Eqs. (8a, b), with replacement of $a_2$ by $a_2^*$, around $\left(N, P, a_2^*\right) = \left(N_e(x), P_e, a_2\right)$ are given as follows:

$$\frac{\partial \Delta N}{\partial t} = \frac{\partial^2 \Delta P}{\partial x^2}, \qquad (23a)$$

$$\frac{\partial \Delta P}{\partial t} = (D_u + D_v)\frac{\partial \Delta N}{\partial t} - D_u D_v \frac{\partial^2 \Delta N}{\partial x^2} + (D_v - D_u)\left(h_N \Delta N + h_P \Delta P + h_a \Delta a_2^*\right), \qquad (23b)$$

where $\qquad h_N(x) = \partial f^*(N_e(x), P_e, a_2)/\partial N \qquad$, $\qquad h_P(x) = \partial f^*(N_e(x), P_e, a_2)/\partial P \qquad$, $h_a(x) = \partial f^*(N_e(x), P_e, a_2)/\partial a_2^*$, $\Delta N = 2\Delta x_p b N_0 \tanh(bx)\operatorname{sech}^2(bx) + n_\varepsilon(x)$, $\Delta P = p_\varepsilon(x)$, and $\Delta a_2^* = \varepsilon a_\varepsilon(x)$. Equation (23a) under the periodic boundary condition leads to $p_\varepsilon$ as follows:

$$p_\varepsilon(x) = -v_p \frac{N_0}{b}\left[\tanh(bx) - \frac{2}{L}x + C_3\right], \qquad (24)$$

where $v_p$ represents the derivative of $\Delta x_p$ with respect to time, that is, $v_p = d\Delta x_p / dt$. By substituting Eq. (24) into Eq. (23b), we obtain the following:

$$\frac{d^2 n_\varepsilon}{dx^2} = 4b^2 \left[1 - 3\operatorname{sech}^2(bx)\right] n_\varepsilon + v_p b N_0 G_n(x) + \varepsilon G_a(x), \qquad (25)$$

where $G_n(x)$ and $G_a(x)$ are defined by:

$$G_n = \frac{2(D_u + D_v)}{D_u D_v} \tanh(bx)\operatorname{sech}^2(bx) + \frac{6N_0}{P_e} \operatorname{sech}^4\left[\tanh(bx) - \frac{2}{L}x + C_3\right], \qquad (26a)$$

$$G_a = \frac{4b^2 N_0}{a_2} \operatorname{sech}^2(bx) a_\varepsilon(x). \qquad (26b)$$

By solving Eq. (25), we obtain $n_\varepsilon$:

$$n_\varepsilon(x) = W_2 \int W_1 \left[v_p b N_0\, G_n + \varepsilon\, G_a\right] dx - W_1 \int W_2 \left[v_p b N_0\, G_n + \varepsilon\, G_a\right] dx, \qquad (27)$$

where $W_1(x)$ and $W_2(x)$ are defined by

$$W_1 = -2\tanh(bx)\operatorname{sech}^2(bx), \qquad (28a)$$

$$W_2 = \frac{-1}{16b} \left\{2\cosh^2(bx) + 5\left[1 - 3\operatorname{sech}^2(bx)\right] + 15bx \tanh(bx)\operatorname{sech}^2(bx)\right\}. \qquad (28b)$$

Considering the periodic boundary condition for $n_\varepsilon(x)$, we obtain the following equation for sufficiently large $L$ (solvable condition):

$$n_\varepsilon(L/2) - n_\varepsilon(-L/2) = W_2(L/2)\int_{-L/2}^{L/2} W_1\left[v_p b N_0 G_n + \varepsilon G_a\right]dx = 0 \ . \qquad (29)$$

This leads to the velocity of the peak:

$$v_p = -\left(\int_{-L/2}^{L/2} W_1 \varepsilon G_a \, dx\right)/b N_0 Z \ , \qquad (30)$$

where $Z$ is given as follows:

$$Z = \int_{-L/2}^{L/2} W_1 G_n \, dx = -\frac{64}{15}b\left[\frac{D_v + D_u}{D_v - D_u}\frac{1}{a_1 a_2} + \frac{L^2 \overline{N}^2}{6 D_v a_2}\left(\frac{3}{7} - \frac{1}{bL}\right)\right]. \qquad (31)$$

Equation (30) indicates that the velocity of the peak is determined by the integral of $W_1 \varepsilon G_a$ with respect to $x$. Taking into consideration that $\varepsilon G_a$ represents the position-dependence of $a_2^*$ and that the position-dependence at the site where the value of $W_1$ is trivial has little influence on the velocity, we can regard $W_1$ as a "sensing window." As shown in Fig. 4, this window has significant value only at the site of the concentration peak. Note that the integral of $W_1 \varepsilon G_a$ is zero when $a_\varepsilon(x)$ is an even function; therefore, the sensing window can detect a gradient (or slope) of $a_2^*$.

Based on this analysis, the existent peak can detect the position-dependence of a parameter and move depending on the gradient at the site of the peak. This property can be called

"localized sensitivity."

## 4.4 Verification of analysis by computations

### 4.4.1 Approximation of one-peak solution

First, we verify the analysis in Section 4.1. We obtain analytical approximations of the one-peak solution as Eqs. (11a, b), and we compare this approximation with the final profile of the numerical simulation of Model II. The computation was performed by setting $L = 10$ and $D_v = 1$ and taking the initial state as $u = 1$ and $v = 1$. The final profile ($t = 200$) is shown in the left panel of Fig. 5A. The solid line indicates the profile of $N$, and the dashed line indicates $P$. The right panel of Fig. 5A shows the approximations given by Eqs. (11a, b), taking the center of the peak as $x_p = -2.15$, which was chosen based on least-square methods. The results of the computations show a good agreement to our analytical results. The approximations were also sufficient when we set $L = 20$, 40, and 80 (data not shown).

### 4.4.2 Instability of two-peak solution

Next, we verify the analysis in Section 4.2.2. According to our analysis, a two-peak solution is unstable and perturbations will grow exponentially with a growth rate $\mu_{anl}$ given by Eq. (21b) with $n = 2$ and $k = 1$, that is, $\mu_{anl} = \dfrac{96 D_v a_2}{L^3 N^2 b}$.

We performed the following experimental computations. First, we obtained a stable one-peak pattern in Model II (Eqs. 3) by taking the size to be $L/2$, where $L = 20$, 30, 40, and taking the initial state as $u = 1$ and $v = 1$. Because we applied the periodic boundary condition to this system, we could set the center of the concentration peak at $x = 0$ by translation. Next, by duplicating this profile ($L/2$) and connecting them, we obtained a new profile ($L$) with two peaks.

We used this profile (*L*) with small perturbations (+0.01*Rnd*) as the initial state of the following simulation. All trials ($D_v$ = 1, 2 and *L* = 20, 30, 40) showed instabilities of two-peak profiles, and we obtained the growth rate, $\mu_{sml}$, from the change in the height of the peak. We logarithmically plotted the growth rates estimated by analytical results ($\mu_{anl}$; Fig. 5B, solid and dashed lines) and those obtained by numerical simulations ($\mu_{sml}$; Fig. 5B, filled circles and squares) against the system size *L*. The results of the analyses and computations were nearly identical.

### 4.4.3 Movement of an existent peak in response to the parameter gradient

Finally, we verify the analysis in Section 4.3. According to our analysis, the existent peak will move when a parameter gains position-dependence. A concentration peak formed in Model II (Eqs. 3) with uniform $a_2$ will move when $a_2$ is replaced by $a_2^* = a_2 \left[ 1 + \dfrac{\varepsilon}{2} \sin\left( 2\pi \dfrac{x}{L} \right) \right]$ with a velocity $v_{anl}$ obtained by Eq. (30) as:

$$v_{anl} = -\frac{\displaystyle\int_{-L/2}^{L/2} \left[ \tanh(bx)\,\mathrm{sech}^4(bx)\sin(2\pi x / L) \right] dx}{\dfrac{16}{15} \left[ \dfrac{D_v + D_u}{D_v - D_u} \dfrac{1}{a_1 a_2} + \dfrac{L^2 \overline{N}^2}{6 D_v a_2} \left( \dfrac{3}{7} - \dfrac{1}{bL} \right) \right]} \varepsilon \ . \tag{32}$$

For the experimental computations, we obtained a stable one-peak pattern in Model II (Eqs. 3) by setting *L* = 10 and taking the initial state as *u* = 1 and *v* = 1. We set the center of the concentration peak at *x* = 0 by translation. Next, we substituted $a_2^*$ for $a_2$ = 2.2 in Eqs. (3). Here the difference between the largest and smallest points is $100\varepsilon$ (%). All trials ($D_v$ = 1, 2 and $\varepsilon$ = 0.02, 0.04, 0.06) showed movement of the existent peaks, and we obtained the velocities, $v_{sml}$, from the results. We plotted the velocities estimated by analytical results ($v_{anl}$; Fig. 5C, solid and dashed lines) and those obtained by numerical simulations ($v_{sml}$; Fig. 5C, filled circles and squares) against the parameter gradients ε. The results of the analyses and computations were nearly identical.

## 5. Rho GTPases model

The models discussed in the previous sections have two components. When a molecule has two states, such as an active and an inactive state, they can be treated as components of a mass conservation system.

The properties of mass conservation systems are also observed in extended mass conservation systems having more than two components. Therefore, we can construct a model involving multiple molecules that satisfy the following conditions: (1) a molecule (X) has two states (Xm and Xc); (2) the total amount of X is conserved; and (3) the diffusion coefficient of Xc is larger than that of Xm.

Here, we propose a model for cell polarity of neutrophils involving the cross-talk of the Rho family of small GTPases (Rac, Cdc42, RhoA). Rho GTPases exhibit guanine nucleotide-binding activity and function as molecular switches, cycling between an inactive GDP-bound state and an active GTP-bound state. In addition, molecules in an active state are located in the plasma membrane, and those in an inactive state are in the cytosol (Fig. 6A; Kaibuchi et al., 1999). It is likely that molecules in the cytosol have larger diffusion coefficients than those in the plasma membrane. Rho GTPases have been reported to interact with one another, and their cross-talk can generate temporal or spatial patterns (Sakumura et al., 2005). According to previous studies, Cdc42 activates Rac (Giniger, 2002; Lim et al., 1996; Nobes and Hall, 1995), and RhoA has mutual inhibitory interactions with Cdc42 and Rac (Giniger, 2002; Leeuwen et al., 1997; Rottner et al., 1999; Sander et al., 1999; Wang et al., 2003). In addition, Rac plays a dominant role in a positive feedback loop, which involves phosphoinositide 3-kinase (PI3K), phosphatidylinositol 3,4,5-triphosphate (PIP$_3$), and F-actin (Li et al., 2003; Srinivasan et al., 2003; Wang et al., 2002;

Weiner et al., 2002).

We construct a model composed of Rho GTPases based on the inferences above (Fig. 6B). We assume that molecules of Rac, Cdc42, and RhoA are activated by guanine nucleotide exchange factors (GEFs; $ka_i$) and are inactivated by GTPase-activating proteins (GAPs; $ki_i$) and that interactions between molecules ($k_{ij}$) are additive to GEFs or GAPs. Some molecule-molecule interactions are stimulation-dependent. Activations of molecules by the stimulation ($ks_i$) are also assumed to be additive to GEFs. The model is as follows:

$$\frac{\partial Rac_m}{\partial t} = Dm_1 \frac{\partial^2 Rac_m}{\partial x^2} - (k_{13}Rho_m + ki_1)Rac_m + (k_{11}Rac_mS + k_{12}Cdc_m + ks_1S + ka_1)Rac_c ,$$

$$\frac{\partial Rac_c}{\partial t} = Dc_1 \frac{\partial^2 Rac_c}{\partial x^2} + (k_{13}Rho_m + ki_1)Rac_m - (k_{11}Rac_mS + k_{12}Cdc_m + ks_1S + ka_1)Rac_c ,$$

$$\frac{\partial Cdc_m}{\partial t} = Dm_2 \frac{\partial^2 Cdc_m}{\partial x^2} - (k_{23}Rho_m + ki_2)Cdc_m + (ks_2S + ka_2)Cdc_c ,$$

$$\frac{\partial Cdc_c}{\partial t} = Dc_2 \frac{\partial^2 Cdc_c}{\partial x^2} + (k_{23}Rho_m + ki_2)Cdc_m - (ks_2S + ka_2)Cdc_c ,$$

$$\frac{\partial Rho_m}{\partial t} = Dm_3 \frac{\partial^2 Rho_m}{\partial x^2} - (k_{31}Rac_m + k_{32}Cdc_m + ki_3)Rho_m + (ks_3S + ka_3)Rho_c ,$$

$$\frac{\partial Rho_c}{\partial t} = Dc_3 \frac{\partial^2 Rho_c}{\partial x^2} + (k_{31}Rac_m + k_{32}Cdc_m + ki_3)Rho_m - (ks_3S + ka_3)Rho_c ,$$

where *Rac*, *Cdc*, and *Rho* with suffixes *m* and *c* denote the concentrations of Rac, Cdc42, and RhoA in the active state and inactive state, respectively. The numerical suffixes represent the following: 1, Rac; 2, Cdc42; and 3, RhoA. $Dm_i$ and $Dc_i$ denote the diffusion coefficients of molecules in the active state and inactive state, respectively. The position-dependent parameter, *S*, denotes the intensity of stimulation.

Using this model, we performed a numerical simulation. The values of parameters and the initial conditions were set as follows: $L = 10$, $Dm_i = 0.04$, $Dc_i = 3$ ($i = 1, 2, 3$), $ks_1= 1$, $ks_2 = 1$, $ks_3 = 1$, $ka_1 = 0.2$, $ka_2 = 0.2$, $ka_3 = 0.2$, $ki_1 = 0.4$, $ki_2 = 0.2$, $ki_3 = 0.2$, $k_{11} = 4$, $k_{12} = 3$, $k_{13} = 5$, $k_{23} = 6$, $k_{31} = 4$, $k_{32} = 2$; initial state, $Xm(x) = 0.3$, $Xc(x) = 0.7$ ($X=Rac$, *Cdc*, *Rho*). The stimulation, *S*, was given by $S = S_m \left\{ 1 + r \cos\left[2\pi\left(x/L - 0.3\right)\right]\right\}$, $S_m = 0.4$, $r = 0.01$.

The results of the simulation ($t > 150$) are shown in Fig. 6C. The peaks of Rac (solid line) and Cdc42 (dotted line) were observed at the site of highest $S$, and RhoA (dashed line) was separated from Rac and Cdc42. These results were in rough agreement with experimental observations (Li et al., 2003; Li et al., 2005; Ridley, 2001; Ridley et al., 2003; Xu et al., 2003).

## 6. Discussion

### 6.1 Specificity and universality of the mass conservation system

Mass conserved models generate multiple peaks from the homogenous state during the early phase, which is explained by Turing instability. But they exhibit characteristic transitions after initial peaks arise: most peaks get smaller and disappear one after another, and only one peak persists. Why is the behavior of the mass conservation system so different from that of ordinary Turing models?

Consider a reaction-diffusion system with vast size ($L \to \infty$) and interval $I$ [$x_1$, $x_2$] within the system, where $x_1$ and $x_2$ are arbitrary but far apart. Can we predict what will happen to the interval $I$? For an ordinary Turing model, it is known that the linearization analysis around the homogenous solution gives us sufficient information. The mass conservation system is more complex, however, because the behavior differs between the case where the components flow into interval $I$ versus the case where they flow out, and we cannot predict which case will occur. The linearization analysis around the homogenous solution gives us information about only the initial transition. This difference of predictability seems to be fundamentally linked to the different behavior and the specificity of the mass conservation system.

We investigated the final realized states of mass conservation systems. Mass conservation models have multiple stationary states, which are spatially homogenous or periodic, including the one-peak state and multiple-peak states. In Section 4, we showed that the multiple-peak stationary states are unstable for Model II. If the homogenous state and multiple-peak states are unstable, we can expect that the system will finally exhibit a one-peak stationary state.

It may be counter-intuitive that any mass conservation system finally exhibits one-peak pattern.

How general are the characters of the mass conservation system described here? For Model II, the final steady state was a one-peak solution regardless of the system size, even when $L$ was infinitely large. But for Model I, the final steady state had two-peaks when we set $L = 80$ (data not shown). It is likely that some mass conservation models have a maximum size to have unique peak. However, it is important to note that this maximum size is independent of the linearization analysis; it depends not on Turing instability but on some other factors. The conditions for the uniqueness of concentration peak will be elucidated in future analyses. In addition to the two models presented here, we have found other mass conservation models that show similar behaviors, which will be presented in another paper.

## 6.2 Biological meanings of mass conservation

Consider a molecule that satisfies the following conditions: (1) the molecule (X) has two states (Xm and Xc); (2) the total amount of X is conserved; and (3) the diffusion coefficient of Xc is larger than that of Xm. Two states of this molecule can be treated as components of a mass conservation system.

Some kinds of small GTPases, such as those of the Rho family, are known to have two states, an active and an inactive state; molecules in the active state are located in the membrane and those in the inactive state in the cytosol (Kaibuchi et al., 1999). Some enzymes involved in the cell polarity of chemotactic cells, such as PI3K and phosphatase and tensin homologue deleted on chromosome 10 (PTEN), are also reported to show a relationship between their activity and membrane binding (Brock et al., 2003; Funamoto et al., 2002; Huang et al., 2003; Iijima and Devreotes, 2002; Iijima et al., 2004). It is reasonable to suppose that molecules in the cytosol diffuse faster than those in the plasma membrane. Thus, these molecules can be considered as components of mass conservation systems.

Chemotactic cells, such as *Dictyostelium* and neutrophils, polarize within a few minutes (30 sec to 3 min) after they are exposed to chemoattractants (Parent and Devreotes, 1999; Servant et al., 2000; Xu et al., 2003). In contrast, regulation of protein levels involves binding of transcriptional factors to DNA, transcribing of the DNA into messenger RNA (mRNA), translation of the mRNA into proteins, and many other processes. Because it is likely that the time scale of cell polarity is much shorter than that of gene regulation, we can assume that the mass of molecules are constant during the polarization of chemotaxis.

## 6.3 Biological meanings of characteristic properties

It is important for chemotactic cells to have only one front-back axis because multiple fronts would prevent fine migration. Subramanian and Narang (2004) referred to this property as "unique localization" and investigated the response of their model to two unequal stimulations. They showed that only one of the two peaks that arise persists, in agreement with our results. It is interesting that their model also contains conserved mass (see Section 6.4).

It has been reported that the front edge of a migrating cell is more sensitive to the new stimulation gradient than is the back edge (Devreotes and Janetopoulos, 2003; Xu et al., 2003). The localized sensitivity focuses the activity of the actin cytoskeleton at the leading edge, resulting in faster movement toward a chemoattractant source (Devreotes and Janetopoulos, 2003). Few mathematical models or theories, however, have been proposed to explain the localization of sensitivity. Mass conservation models respond to the parameter position-dependence and their sensitivity is localized at the site of the existent peak. Because concentration of chemoattractant can be treated as a position-dependent parameter, these models can explain the localized sensitivity to the chemoattractant gradient.

## 6.4 Multiple-component model for cell polarity

Multiple-component models involving at least two components whose sum is conserved readily exhibit the properties investigated in this paper, such as the uniqueness of axis and localization of sensitivity. The Rho GTPases model is an example of a multiple-component model. Many models have been proposed to explain cell polarity during chemotaxis, and some of them show very similar properties (Narang et al., 2001; Postma and Van Haastert, 2001; Skupsky et al., 2005; Subramanian and Narang, 2004). These models have some factors (or molecules) that have two or more states and whose masses are conserved.

Models of chemotaxis should take into account directional sensing and signal amplification (Parent and Devreotes, 1999). As shown in our simulation results, the mass conservation system with a position-dependent parameter forms a stable pattern with one distinct peak from a homogenous state at the site that is determined by the slight gradient of the parameter value. As for the Rho GTPases model, the peaks of Rac and Cdc42 were observed at the site of the highest $S$, and RhoA was separated from Rac and Cdc42. These results indicate that the mass conservation system can explain directional sensing and signal amplification.

In conclusion, mass conservation reaction-diffusion models show directional sensing, signal amplification, uniqueness of axis, and localization of sensitivity. All of these properties are necessary for cell polarity. Furthermore, our Rho GTPase model has these properties and is consistent with experimental observations. Thus, these properties of mass conservation systems may explain the formation of cell polarity as well.

**Acknowledgments**

We would like to thank Yu-ichi Ozaki, Satoru Sasagawa, Kazuhiro Fujita, and Hidetoshi Urakubo for valuable discussions. We are also grateful to Yuya Terashima and Kouji Matsushima for important comments about chemotaxis of neutrophils. This work was supported by a grant in-aid for scientific research from the Ministry of Education, Culture, Sports, Science and Technology of Japan.

**Appendix**

Note that Eq. (9b) has the same formulation as classical Newton mechanics. We define $V(N_e)$ as

$$V\left(N_e; P_e\right) = -\frac{D_v - D_u}{D_u D_v} \int_0^{N_e} f^*\left(N, P_e\right) dN \, , \tag{A.1}$$

and Eq. (9b) implies

$$dx = \frac{dN_e}{\sqrt{2\left[E - V\left(N_e; P_e\right)\right]}} \, , \tag{A.2}$$

where $E$ is a constant value, corresponding to period and total mass of $N_e(x)$. The period $\lambda$ and the average mass $\overline{N} = \frac{1}{\lambda} \int_0^\lambda N_e\left(x\right) dx$ satisfy the following equations:

$$\lambda = 2\int_{N_{\min}}^{N_{\max}} \frac{dN_e}{\sqrt{2\left[E - V\left(N_e; P_e\right)\right]}} \, , \tag{A.3}$$

$$\overline{N} = \frac{2}{\lambda} \int_{N_{\min}}^{N_{\max}} \frac{N_e \, dN_e}{\sqrt{2\left[E - V\left(N_e; P_e\right)\right]}} \, , \tag{A.4}$$

where $N_{\min}$ and $N_{\max}$ are minimum and maximum levels of $N_e(x)$, respectively, and are derived from $V\left(N_{\min}\right) = V\left(N_{\max}\right) = E$ ( $0 < N_{\min} < \overline{N} < N_{\max}$ ). Note that Eqs. (A.3) and (A.4) give the relationship among $P_e$, $\lambda$, and $\overline{N}$, where $\overline{N}$ is derived from the initial condition of ($u$, $v$) straightforwardly. For Model II (Eqs. 3),

$$V\left(N_e; P_e\right) = \frac{D_v - D_u}{D_u D_v} \frac{a_1 P_e}{3 D_v} \left( N_e^{\ 3} - \frac{3 D_v a_2}{2 P_e} N_e^{\ 2} \right), \tag{A.5}$$

and $E$ can range between $E_* < E < 0$ for $N_e(x)$ to be a periodic solution. Here $E_* = -\frac{D_v - D_u}{D_u D_v} \frac{D_v^{\ 2} a_1 a_2^{\ 3}}{6 P_e^{\ 2}}$. As $E$ becomes smaller ( $E \to E_*$ ), the period $\lambda$ converges to $\lambda_{\min}$, which is the shortest wavelength in the periodic solutions. As $E$ becomes larger ($E \to 0$), the period $\lambda$ diverges. The solution of $N_e(x)$ at $E = 0$ corresponds to the separatrix of Eq. (9b), indicating infinite period ($\lambda \to \infty$). The explicit form of $N_e(x)$ for $E = 0$ can be obtained by

$$N_e(x) = \frac{3D_v a_2}{2P_e} \operatorname{sech}^2\left[\frac{1}{2}\sqrt{\frac{D_v - D_u}{D_u D_v} a_1 a_2}\left(x - x_p\right)\right], \qquad (A.6)$$

which has the sole peak at $x = x_p$ and decays to zero as $x \to \pm\infty$. For a sufficiently large system, Eq. (A.6) is a good approximation of the solution for $-L/2 < x < L/2$. Equations (A.6) and (10) lead to Eqs. (11a, b) and (12a–c).

**Figure legends**

**FIGURE 1.** Transient behavior and final stable states of the system. Using Model I, we performed experimental simulations to show the transient behavior and instability of multiple peaks. (A) Behavior of the system with $L = 10$ from the homogenous stationary state with small perturbations. (B) Behavior of the system with $L = 20$ from the homogenous stationary state with small perturbations. (C) Transitions of four peaks that arose in the system with $L = 20$ (see panel B). (D) Behavior of the system with $L = 10$ from the two-peak state with small perturbations. All vertical axes indicate the levels of $u$.

**FIGURE 2.** Behavior of the system including a position-dependent parameter. Using Model I, we performed experimental simulations to investigate the dynamics of the concentration peak for a position-dependent parameter. (A) Behavior of the system from the homogenous stationary state. We substituted $a_3^* = 2 + 0.06 \sin\left(2\pi \dfrac{x}{L}\right)$ for $a_3 = 2$ in Eqs. (2). (B) Movement of existent peak depending on the position-dependence. (C) Movement of existent peak when there is overlap between the peak and the local position-dependence of $a_3^*$. The arrow indicates the direction of movement. The dashed line indicates the spatial profile of $a_3^*$ (magnified in the figure). (D) Movement of existent peak when there is little overlap between the peak and the local position-dependence of $a_3^*$. All vertical axes indicate the level of $u$.

**FIGURE 3.** Stability analysis of multiple-peak solution. We seek growable perturbations $\left(n_\mu, p_\mu\right)$ for the periodic solutions as discussed in the text. First, we show the perturbations for the one-peak solution, without consideration of boundary conditions: (A) $n_\mu(x)$; and (B) $p_\mu(x)$. The dashed line with an arrow in panel B indicates the approximation to a piecewise linear function. Next, we show the perturbations for multiple-peak solutions. We can describe a perturbation by a set of $p_j$, where $p_j$ is the value of $p_\mu$ at the center of the $j$th peak. As shown in the text, we obtain $p_j$ as $p_j{}^{(k)} = \cos\left(\dfrac{2k\pi}{n} j + \theta_0\right)$, where $k$ is an integer ( $1 \le k \le n/2$ ), corresponding to a mode of perturbation, and $\theta_0$ is an arbitrary constant. Panels C, D, E, and F show the most growable perturbations for two-, three-, four-, and five-peak solution, respectively. For each $n$, the mode of perturbation, $k$, that gives the largest value to $\mu$ is determined by

$$\mu^{(k)} = \frac{2n^3 P_e b}{N_0 L} \sin^2\left(\frac{k\pi}{n}\right).$$

**FIGURE 4.** Sensing window of the existent peak that detects the position-dependence of a parameter. The velocity of the existent peak responding to a parameter gradient is determined by $\int_{-L/2}^{L/2} W_1(x)\varepsilon Ga(x)\,dx$, where $\varepsilon G_a(x)$ represents the position-dependence of the parameter. Here we can regard $W_1(x)$ as a "sensing window." The upper panel shows the profile of the existent peak, $N(x)$; the lower panel indicates $W_1(x)$.

**FIGURE 5.** Verification of analysis by computations. (A) Approximation of one-peak solution. The left panel indicates the result of numerical simulation (Model II); the right panel indicates the analytical approximation. (B) Instability of two-peak solution. A two-peak state is unstable in Model II and some perturbation grows. We compare the growth rates estimated by analysis with those obtained by simulations. For simulations, we varied $D_v$ (= 1 or 2) and $L$ (= 20, 30, or 40);

thus, six trials were performed. The axes indicate μ and $L$ in double logarithmic scales. (C) Movement of existent peak is dependent on the parameter gradient. An existent peak moves when a gradient is given to the parameter. We compare the velocity estimated by analysis with those obtained by simulations. For simulations, we varied $D_v$ (= 1 or 2) and ε (= 0.02, 0.04, or 0.06); thus, six trials were performed.

**FIGURE 6.** Rho GTPases model. (A) The Rho family of GTPases, which are localized in the membrane or cytosol, have conserved mass and show slower diffusion in the membrane than in the cytosol. (B) The diagram of the model with Rho GTPases (details are in the text). (C) Spatial profiles of Rac (solid), Cdc42 (dashed), and RhoA (dotted) after the steady state is achieved. The stimulation with a maximum point at $x/L$ = 0.3 (2% gradient) is given. The thin line indicates the spatial profile of the stimulation (magnified in the figure).

**FIGURE 1**

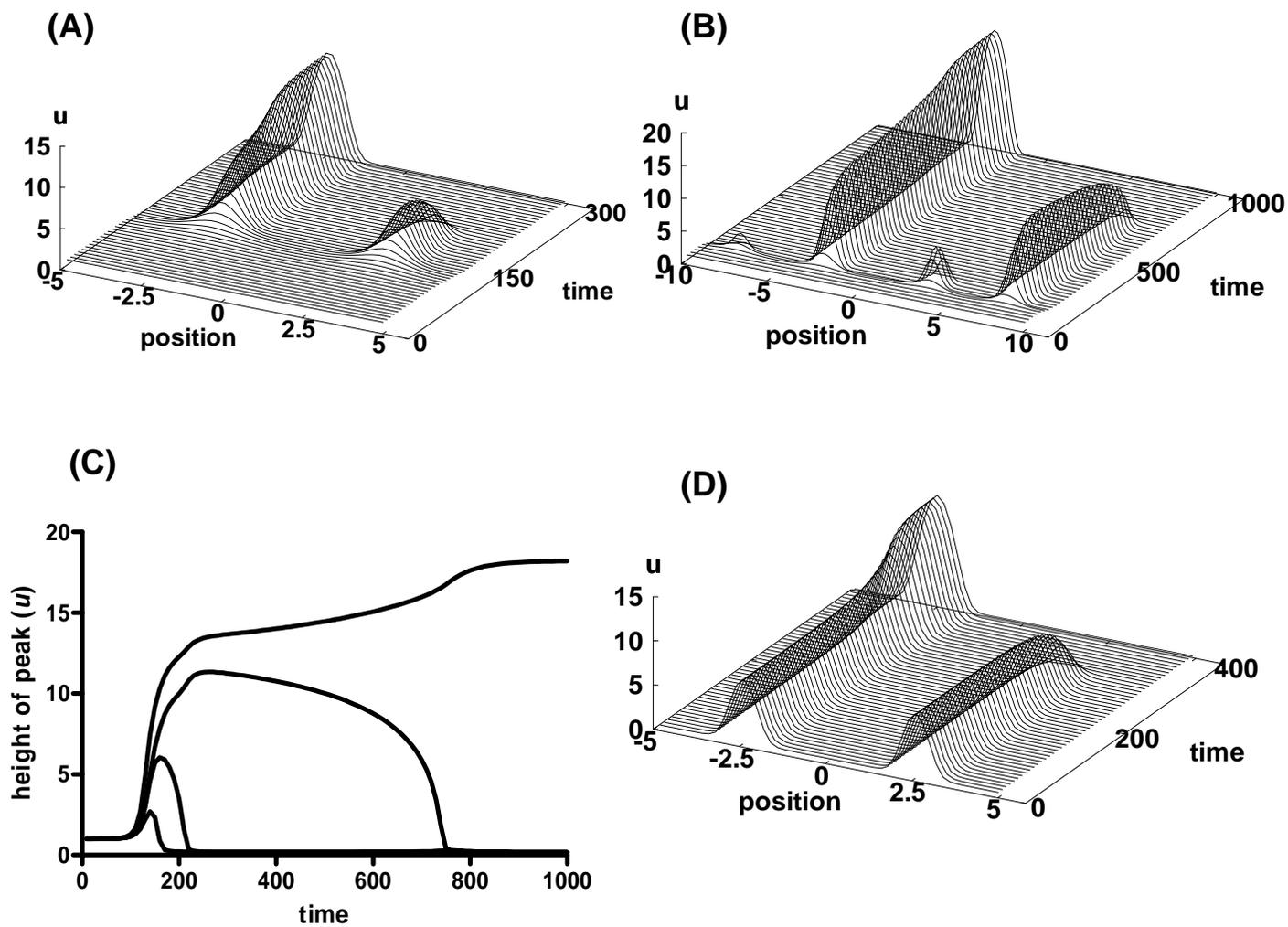

**FIGURE 2**

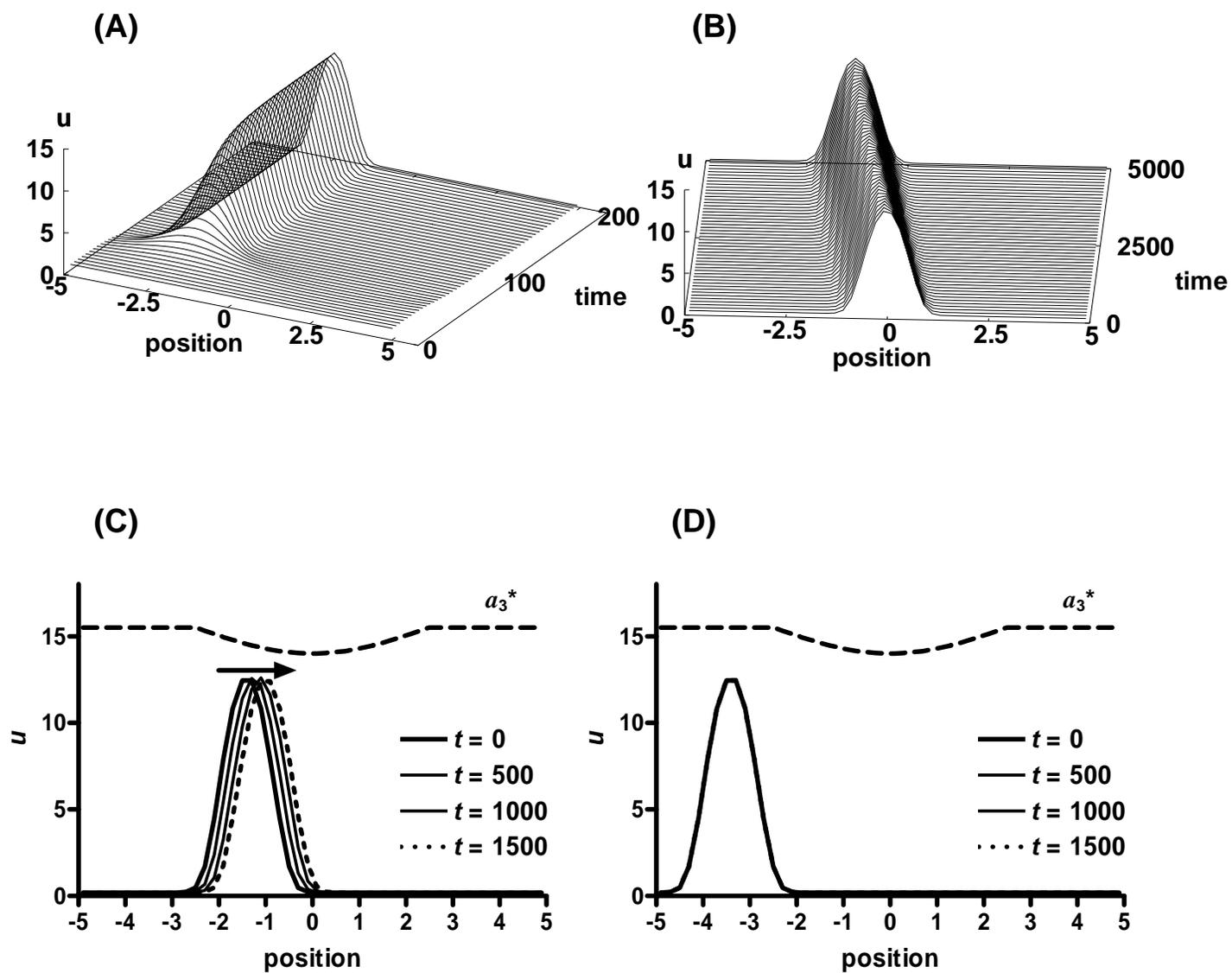

**FIGURE 3**

**(A)**

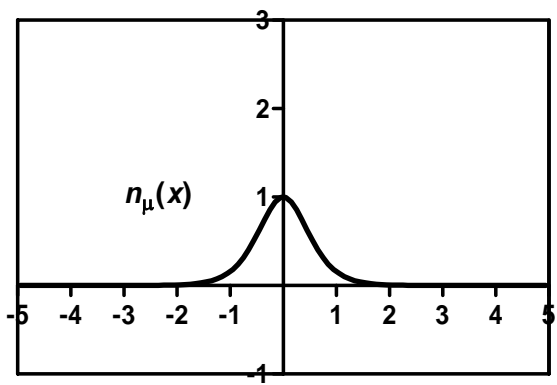

**(B)**

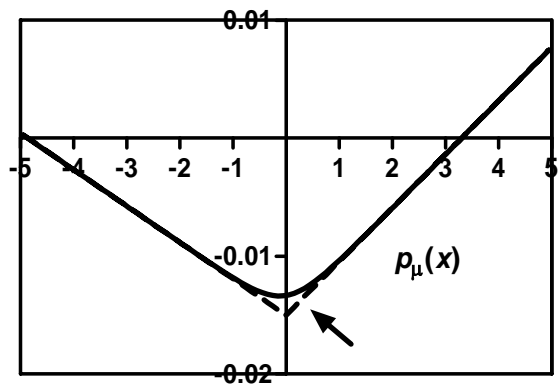

**(C)** $n = 2, k = 1, \theta_0 = 0$

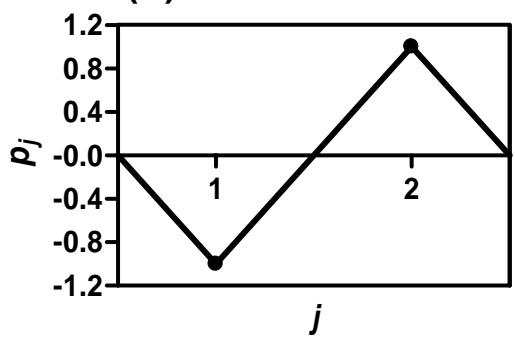

**(D)** $n = 3, k = 1, \theta_0 = \frac{2\pi}{3}$

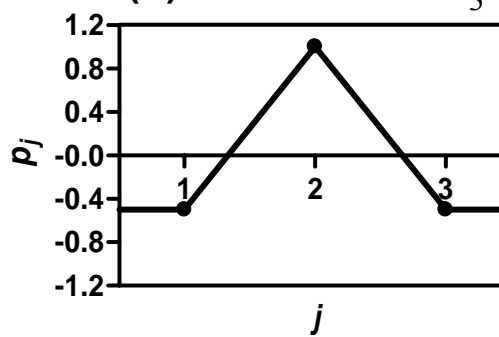

**(E)** $n = 4, k = 2, \theta_0 = 0$

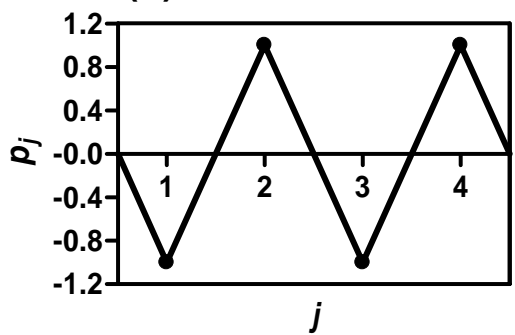

**(F)** $n = 5, k = 2, \theta_0 = \frac{8\pi}{5}$

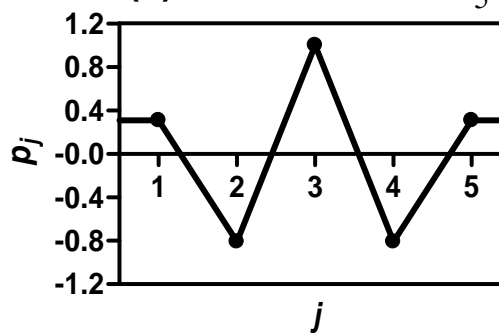

**FIGURE 4**

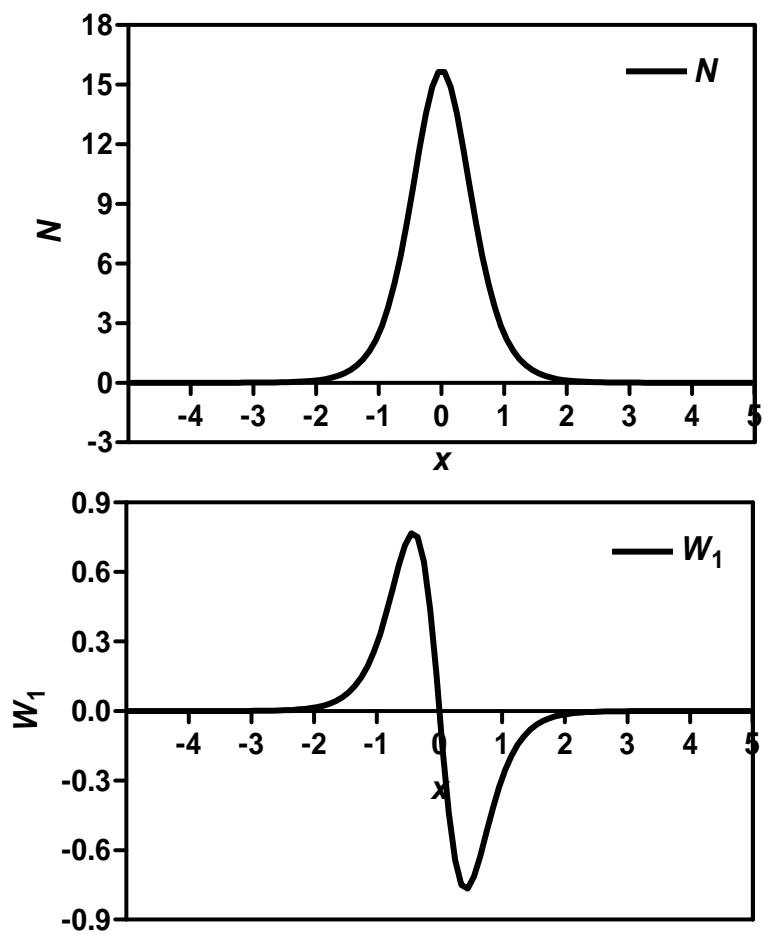

**FIGURE 5**

**(A)**

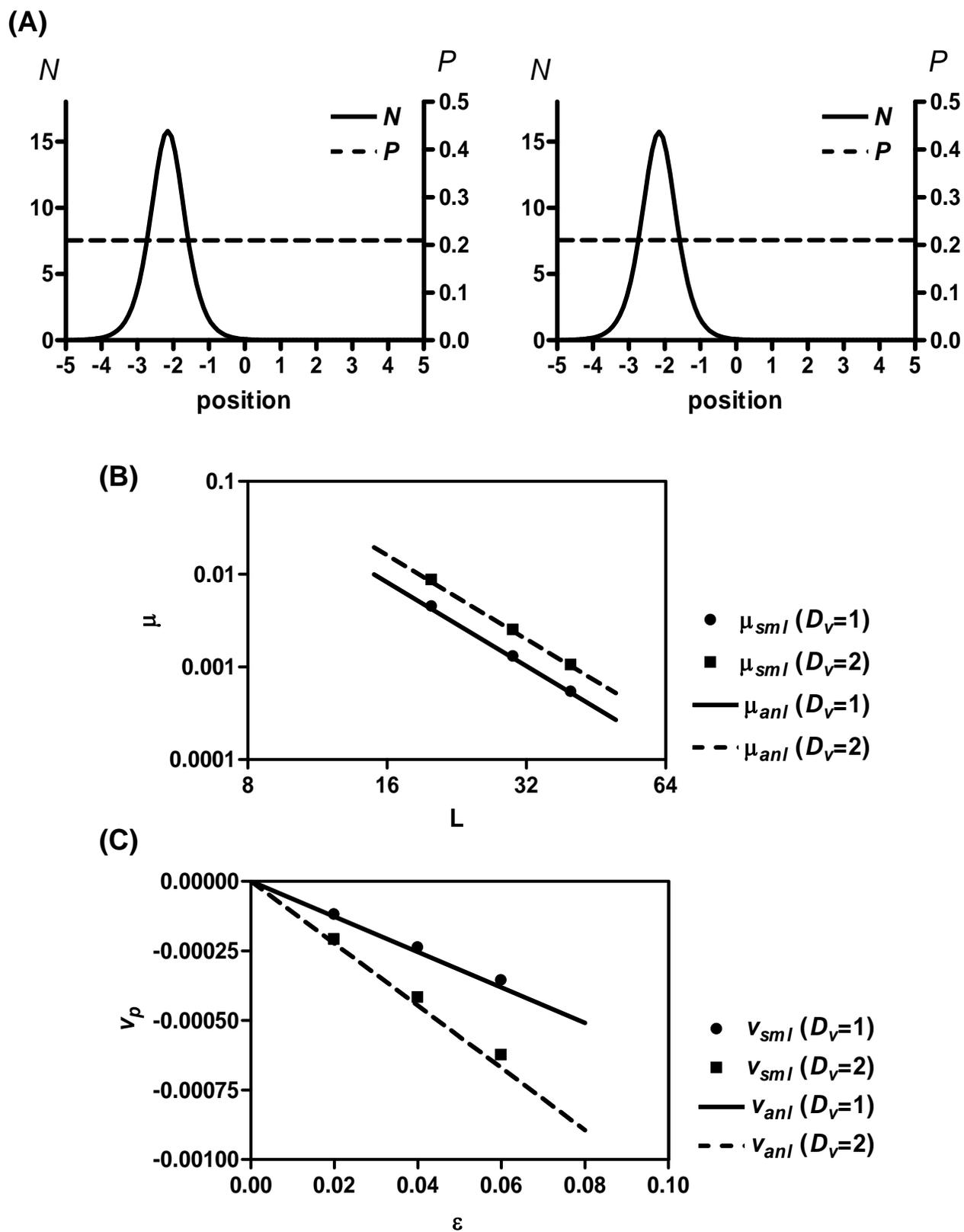

**FIGURE 6**

**(A)**

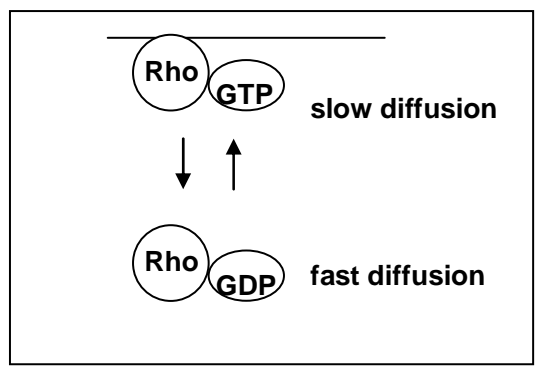

**(B)**

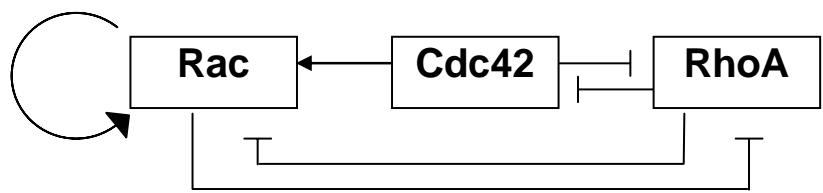

**(C)**

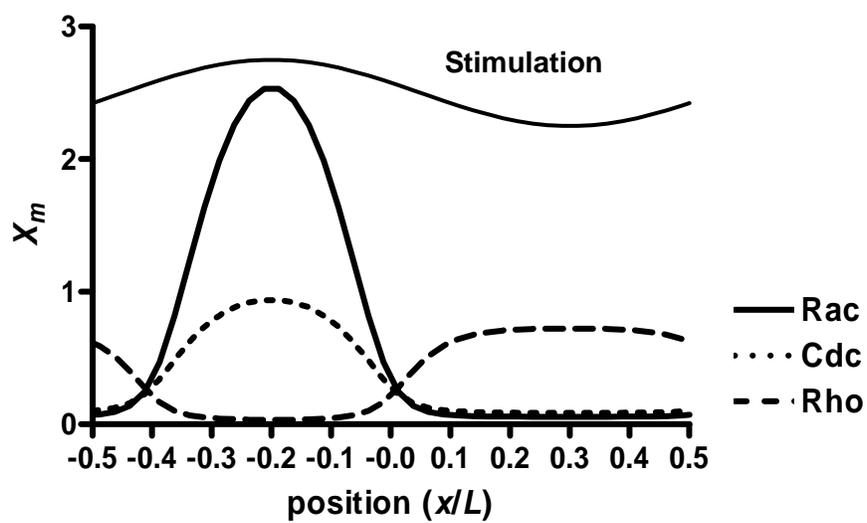